\documentclass[fleqn,10pt]{wlscirep}
\usepackage[utf8]{inputenc}
\usepackage[T1]{fontenc}
\usepackage{hyperref}

\usepackage{bm}
\title{Global Short-Term Forecasting of Covid-19 Cases}

\author[1,+]{Thiago de Paula Oliveira}
\author[2,+,*]{Rafael de Andrade Moral}
\affil[1]{National University of Ireland Galway, Insight Centre for Data Analytics and School of Mathematics, Statistics, and Applied Mathematics, Galway, H91 TK33, Ireland}
\affil[2]{Maynooth University, Department of Mathematics and Statistics and Hamilton Institute, Maynooth, W23 F2H6, Ireland}

\affil[*]{rafael.deandrademoral@mu.ie}

\affil[+]{these authors contributed equally to this work}

\keywords{Bayesian hierarchical models; Coronavirus pandemic; Lockdown policies; State-space models}

\begin{abstract}
The continuously growing number of COVID-19 cases pressures healthcare services worldwide \cite{Kassem2020}. Accurate short-term forecasting is thus vital to support country-level policy making \cite{Goodell2020}. The strategies adopted by countries to combat the pandemic vary, generating different uncertainty levels about the actual number of cases \cite{He2020, Kassem2020}. Accounting for the hierarchical structure of the data and accommodating extra-variability is therefore fundamental. We introduce a new modelling framework to describe the course of the pandemic with great accuracy, and provide short-term daily forecasts for every country in the world. We show that our model generates highly accurate forecasts up to six days ahead, and use estimated model components to cluster countries based on recent events. We introduce statistical novelty in terms of modelling the autoregressive parameter as a function of time, increasing predictive power and flexibility to adapt to each country. Our model can also be used to forecast the number of deaths, study the effects of covariates (such as lockdown policies), and generate forecasts for smaller regions within countries. Consequently, it has strong implications for global planning and decision making. We constantly update forecasts and make all results freely available to any country in the world through an online Shiny dashboard.
\end{abstract}
\begin{document}

\flushbottom
\maketitle

\thispagestyle{empty}

\section*{Introduction}

Outbreaks of the COVID-19 pandemic have been causing worldwide socioeconomic and health concerns since December 2019, putting high pressure on healthcare services. SARS-CoV-2, the causative agent of COVID-19, spreads efficiently and, consequently, the effectiveness of control measures depends on the relationship between epidemiological variables, human behaviour, and government intervention to control the spread of the disease \cite{He2020, West2020}. Different attempts to model the virus outbreaks in many countries have been made, involving mechanistic models \cite{Baker2018, Giordano2020}, and extensions based on susceptible-infected-recovered (SIR) systems \cite{Bjornstad2020, Toda2020, Calafiore2020}. Countries have also put together task forces to work with COVID-19 data and their indirect impact on the population, economy, banking and insurance, and financial markets \cite{Altena2020, Cohen2020, Goodell2020, Perkins2020}. In addition, funding agencies have put together rapid response calls worldwide for projects that can help dealing with this pandemic. However, further investment is still needed to foment priority research involving SARS-CoV-2, so as to establish a high level coordination of essential, policy-relevant, social and mental health science \cite{Holmes2020, Wilson2020}.

This pandemic is associated with high basic reproduction numbers \cite{Lai2020, Wang2020}, spreading with great speed since a significant number of infected individuals remain asymptomatic, while still being able to transmit the virus \cite{Muniyappa2020}. In the UK, for example, it is estimated that approximately $4.4\%$ of people who tested positive for SARS-CoV-2 overall required hospitalisation, increasing to $17-27\%$ in the group of persons aged 65 years or older. A pressing concern here is how to avoid bringing healthcare systems to a collapse \cite{Ferguson2020, Muniyappa2020}. Knowing how the outbreak is progressing is crucial to predict whether or when this will happen, and therefore to plan and implement measures to reduce the number of cases so as to avoid it. 

Policies for reducing the number of infected people such as social distancing and movement restrictions have been put in place in many countries, but for many others a full lockdown may be very difficult (if not impossible) to implement. This also depends heavily on the country's political leadership, socioeconomic reality, and epidemic stage \cite{Kassem2020, Mehtar2020}. In this context, accurate short-term forecasting would prove itself invaluable, especially for systems on the brink of collapse and countries whose governments must consider trade-offs between lockdowns and avoiding full economical catastrophe.

The main problem is that not only is this disease new, but there are also many factors acting in concert resulting in a seemingly unpredictable outbreak progression. Forecasting with great accuracy under these circumstances is very difficult. Here we propose a new modelling framework, based on a state-space hierarchical model, that is able to generate forecasts with very good accuracy for up to seven days ahead. To aid policy making and effective implementation of restrictions or reopening measures, we provide all results as an R Shiny Dashboard, including week-long forecasts for every country in the world whose data is collected and made available by the European Centre for Disease Prevention and Control (ECDC).

\section*{Results}

\begin{figure}[ht!]
    \centering
    \includegraphics[width = \textwidth]{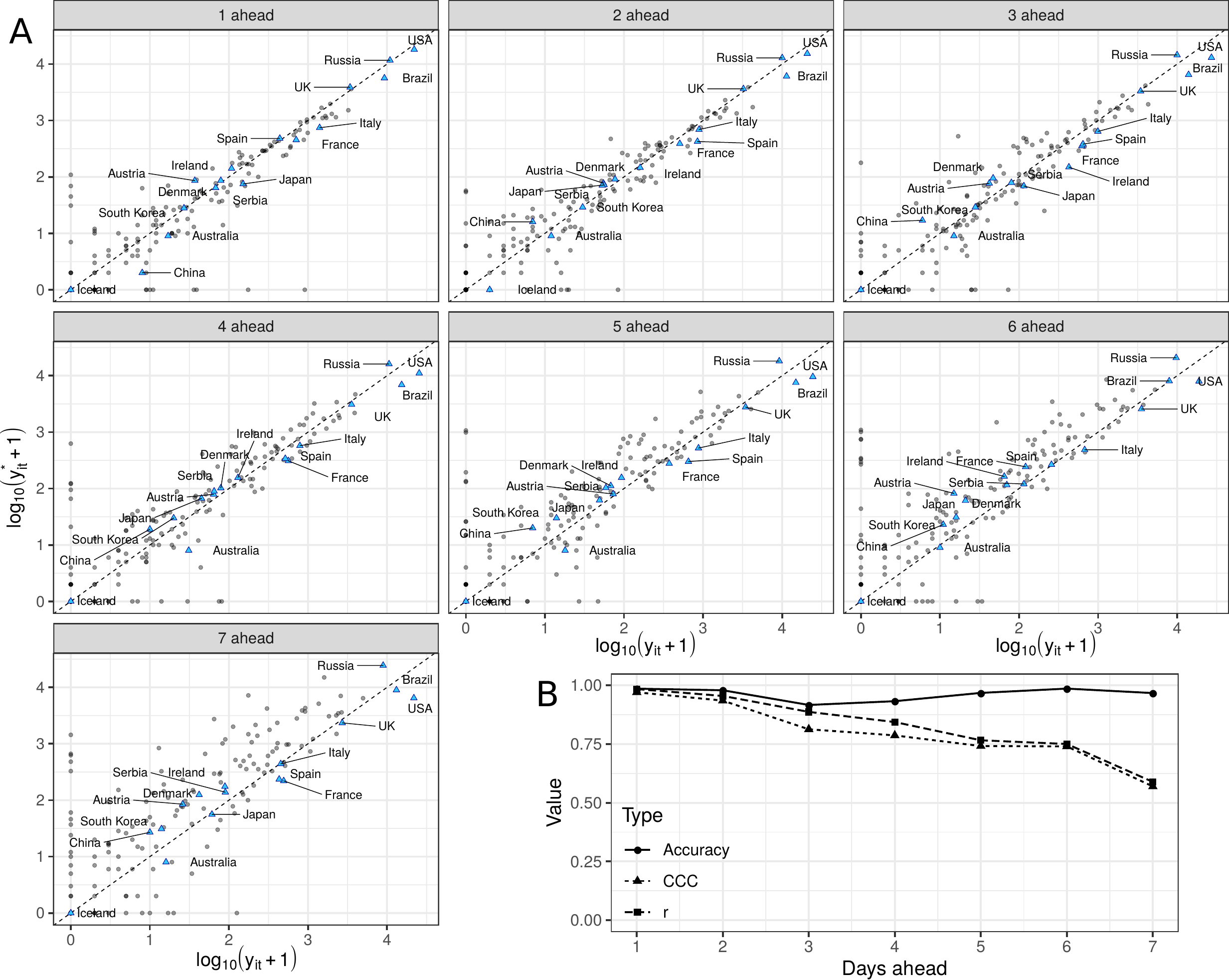}
    \caption{\textbf{A} Logarithm of the observed $y_{it}$ versus the forecasted daily number of cases $y^*_{it}$ for each country, for up to seven days ahead, where each day ahead constitutes one panel. The forecasts were obtained from the autoregressive state-space hierarchical negative binomial model, fitted using data up to 13-May-2020. The first day ahead corresponds to 14-May-2020, and the seventh to 20-May-2020. Each dot represents a country, and the sixteen countries shown in Figure~\ref{fig:gamma} are represented by blue triangles. We add 1 to the values before taking the logarithm. \textbf{B} Observed accuracy, concordance correlation coefficient (CCC) and Pearson correlation ($r$) between observed ($y_{it}$) and forecasted ($y^*_{it}$) values for each of the days ahead of 13-May-2020.}
    \label{fig:forecast}
\end{figure}

Our model displayed an excellent predictive performance for short-term forecasting. We validated the model by fitting it to the data up to 13-May-2020, after removing the last seven time points (from 14-May-2020 until 20-May-2020), and compared the forecasted values with the observed ones (Figure~\ref{fig:forecast}\textbf{A}). We observe that it performs very well, especially for the first six days ahead (Figure~\ref{fig:forecast}\textbf{B}). Even though performance falls substantially for the seventh day ahead, with a concordance correlation between observed and forecasted values close to $0.5$, there are still many countries for which the forecasted daily number of new cases is very close to the observed one. We carried out this same type of validation study using data up to 6-May-2020 and up to 29-Apr-2020, and the results were very similar, with a concordance correlation between observed and forecasted values greater than $0.75$ for up to five days ahead (see Supplementary Materials).

\begin{figure}[ht!]
    \centering
    \includegraphics[width = \textwidth]{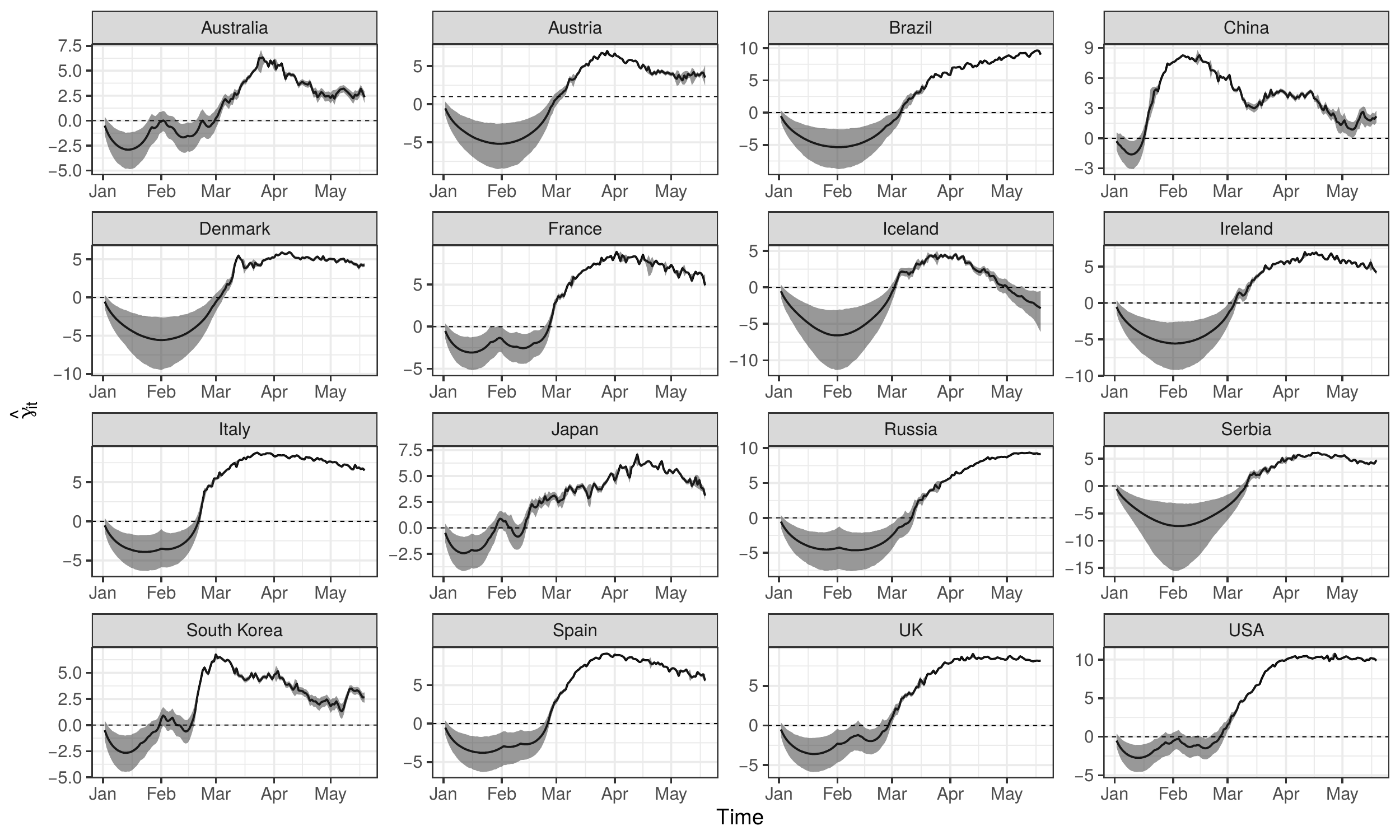}
    \caption{Posterior means of the autoregressive component $\gamma_{it}$ (solid lines) and associated 95\% credible intervals (shaded areas) for each of sixteen selected countries from the pool of 210 countries and territories in the data, from 1-Jan-2020 until 20-May-2020.}
    \label{fig:gamma}
\end{figure}

The autoregressive component in the model has a direct relationship with the pandemic behaviour over time for each country (see Supplementary Materials). It is directly proportional to the natural logarithm of the daily number of cases, given what happened in the previous day. Therefore, it is sensitive to changes and can be helpful detecting a possible second wave. See, for example, its behaviour for Ireland, Spain, Italy and France -- it shows that the outbreak is decaying, however it may still take time to subside completely (Figure~\ref{fig:gamma}). On the other hand, in Iceland the outbreak has ended, according to the estimated autoregressive component. In China and South Korea, however, even though it appears that the outbreak has come to an end, our results show that a second outbreak could begin, and hence countries must be very cautious when relaxing restrictions. In the UK and the United States, it seems as if the outbreak is taking a long time to subside, and in Brazil and Russia it has not even reached its peak yet.

The estimates for the model parameters suggest that about 10\% of the number of reported cases can be viewed as contributing to extra variability, and possibly may consist of outlying observations (see the model estimates in the Supplementary Materials). These observations may be actual outliers, but it is likely that this is a feature of the data collection process. In many countries, the data that is recorded for a particular day actually reflects tests that were done over the previous week or even earlier. This generates aggregated-level data, which is prone to exhibiting overdispersion, which is accounted for in our model, but for some observations this variability is even greater, since they reflect a large number of accumulated suspected cases that were then confirmed. There is also a large variability between countries in terms of their underlying autoregressive processes (see the estimates for the variance components in the Supplementary Materials). This corroborates the fact that countries are dealing with the pandemic in different ways, and also may collect and/or report data differently.

\begin{figure}[ht!]
    \centering
    \includegraphics[width = \textwidth]{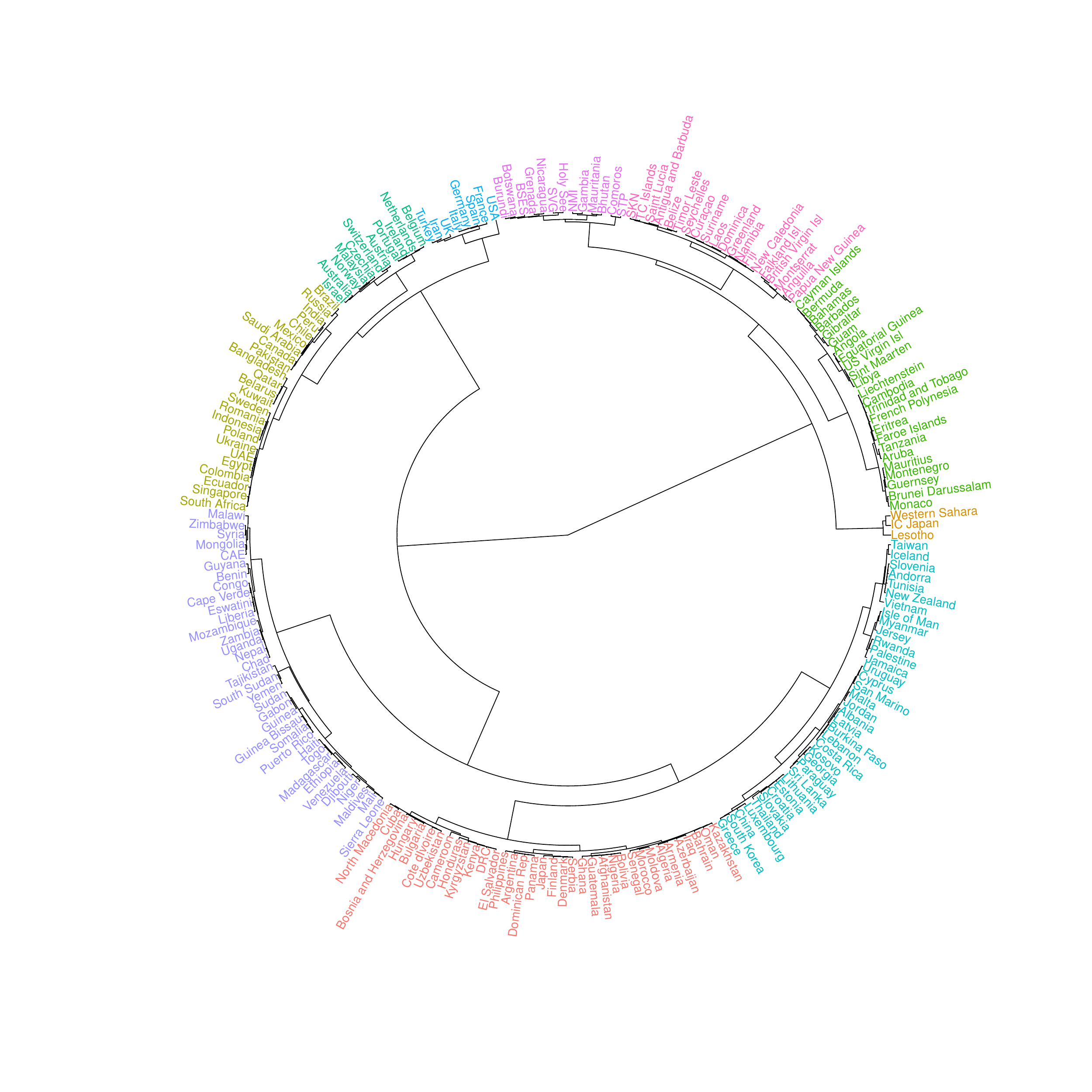}
    \caption{Dendrogram representing the hierarchical clustering of countries based on their estimated autoregressive parameters $\hat\gamma_{it}$ from 20-Mar-2020 to 19-May-2020. The clustering used Ward's method and pairwise dynamic time warp distances between the countries' time series. Each of 10 clusters is represented with a different colour. Country abbreviations: BSES = Bonaire, Saint Eustatius and Saba; IC Japan = Cases on an international conveyance - Japan; CAE = Central African Republic; DRC = Democratic Republic of the Congo; NMI = Northern Mariana Islands; SKN = Saint Kitts and Nevis; SVG = Saint Vincent and the Grenadines; STP = S\~ao Tom\'e and Pr\'incipe; TC Islands = Turks and Caicos Islands; UAE = United Arab Emirates.}
    \label{fig:cluster}
\end{figure}

We propose clustering the countries based on the behaviour of their estimated autoregressive parameter over the last 60 days (Figure~\ref{fig:cluster}). This gives governments the opportunity to see which countries have had the most similar recent behaviour of the outbreak, and study similar or different measures taken by these other countries that may help determine policy. We observe, for example, that Brazil and Russia have been experiencing a very similar situation recently; the same can be said about France, Germany, Italy and Spain. Our R Shiny dashboard displays up-to-date results in terms of forecasts and also country clustering, and can be accessed at \href{https://prof-thiagooliveira.shinyapps.io/COVIDForecast}{https://prof-thiagooliveira.shinyapps.io/COVIDForecast}. Through the dashboard, users can choose to highlight a different number of clusters, which may provide different insights.

\section*{Discussion}

Our modelling framework allows for forecasting the daily number of new COVID-19 cases for each country and territory for which data has been gathered by the ECDC. It introduces statistical novelty in terms of modelling the autoregressive parameter as a function of time. This makes it highly flexible to adapt to changes in the autoregressive structure of the data over time. In the COVID-19 pandemic, this translates directly into improved predictive power in terms of forecasting future numbers of daily cases. Our objective here is to provide a simple, yet not overly simplistic, framework for country-level decision-making, and we understand this might be easier for smaller countries when compared to nations of continental dimensions, where state-level decision-making should be more effective \cite{White2020}. Moreover, the model can be adapted to other types of data, such as number of deaths, and also be used to obtain forecasts for smaller regions within a country. A natural extension would be to include an additional hierarchical structure taking into account the nested relationship between cities, states, countries and ultimately continents, while also accommodating the spatial autocorrelation. This would allow for capturing the extra-variability introduced by aggregating counts over different cities and districts.

We remark that one must be very careful when looking at the forecasted number of cases. These values must not be looked at in isolation. It is imperative that the entire context is looked at and that we understand how the data is actually generated \cite{Vespignani2020}. The model will obtain forecasts based on what is being reported, and therefore will be a direct reflection of the data collection process, be it appropriate or not. When data collection is not trustworthy, model forecasts will not be either. Our estimated number of outlying observations of approximately 10\% is relatively high, and when looking at each country's time series we observe that countries that display unusual behaviour in terms of more outlying observations are usually related to a poorer model predictive performance as well. This suggests that the data collection process is far from optimal. Looking at the full context of the data is therefore key to implementing policy based on model forecasts.

As self-criticism, our model may possibly be simplistic in the sense that it relies on the previous observation to predict the next one, but does not rely on mechanistic biological processes, such as SEIR-type models \cite{Satsuma2004, DeLaSen2015, Baker2018}. These types of models allow for a better understanding of the pandemic in terms of the disease dynamics, and are able to provide long-term estimates \cite{Giordano2020}. However, as highlighted above, our objective is short-term forecasting, which is a task that is already very difficult. Long-term forecasting is even more difficult, and we believe it could even be speculative when dealing with its implications pragmatically. A more tangible solution to this issue is to combine the short-term forecasting power of our proposed modelling framework with the long-term projections provided by mechanistic models, so as to implement policy that can solve pressing problems efficiently, while at the same time looking at how it may affect society in the long run. We acknowledge all models are wrong, including the one presented here. We have showed, however, that it provides very good forecasts for up to a week ahead, and this can help inform on country reopening policies.

Even though the model performs very well, we stress the importance of constantly validating the forecasts, since not only can the underlying process change, but also data collection practices change (to either better or worse). Many countries are still not carrying out enough tests, and hence the true number of infections is sub-notified \cite{Li2020}. This hampers model performance significantly, especially that of biologically realistic models \cite{Baker2018}.

There is a dire need for better quality data \cite{Vespignani2020}. This is a disease with a very long incubation period (up to 24 days\cite{Guan2020}), with median value varying from 4.5 to 5.8 days\cite{Lauer2020}, which makes it even harder to pinpoint exactly when the infection was actually transmitted from one person to the other. Furthermore, around $97.5\%$ of those who develop symptoms will do so in between 8.2 to 15.6 days \cite{Lauer2020}. The number of reported new cases for today reflects a mixture of infections that were transmitted at any point in time over the last few weeks. The time testing takes also influences this. One possible alternative is to model excess deaths when compared to averages over previous years \cite{Kysely2009, Nunes2011}. This is an interesting approach, since it can highlight the effects of the pandemic (see, e.g., the insighful visualisations at \href{https://ourworldindata.org/excess-mortality-covid}{Our World in Data} \cite{Ritchie2020}, and the \href{https://www.ft.com/content/a26fbf7e-48f8-11ea-aeb3-955839e06441}{Finantial Times} \cite{finantial2020}). Again, this is highly dependent on the quality of the data collection process not only for COVID-19 related deaths, but also those arising from different sources. Many different teams of data scientists and statisticians worldwide are developing different approaches to work with COVID-19 data \cite{Calafiore2020, He2020, Petropoulos2020, Wang2020}. It is the duty of each country's government to collect and provide accurate data. This way, it can be used with the objective of improving healthcare systems and general social wellbeing.

The developed R Shiny Dashboard displays seven-day forecasts for all countries and territories whose data are collected by the ECDC, and updated clustering of countries based on the last 60 days. This can help governments currently implementing or lifting restrictions. It is possible to compare government policies between countries at similar stages of the pandemic to determine the most effective courses of action. These can then be tailored by each particular government to their own country. The efficiency of measures put in place in each country can also be studied using our modelling framework, since it easily accommodates covariates in the linear predictor. Then, the contribution of these country-specific effects to the overall number of cases can serve as an indicator of how they may be influencing the behaviour of the outbreaks over time.

Government policies are extremely dependent on the reality of each country. It has become clear that there are countries that are well able to withstand a complete lockdown, whereas others cannot cope with the economic downfall \cite{Mehtar2020}. The issue is not only economic, but also of newly emerging health crises that are not due to COVID-19 lethality alone, but to families relying on day-to-day work to guarantee their food supplies. For these countries, there is a trade-off between avoiding a new evil versus amplifying pre-existing problems or even creating new ones. In such circumstances, it is indeed very difficult to create a one-size-fits-all plan, which makes it even more vital to strive for better data collection practices.

We hope to be able to contribute to the development of efficient short-term response to the pandemic, for countries whose healthcare systems are at capacity, and countries implementing reopening plans. By providing a means of comparing recent behaviour of the outbreak between countries, we also hope to provide a means to opening dialogue between countries going through a similar stage, and those who have gone through similar stages before.

\section*{Methods}

\subsection*{Data acquisition}

The data was obtained from the European Centre for Disease Prevention and Control (ECDC), and the code is implemented such that it downloads and processes the up-to-date data from \href{https://www.ecdc.europa.eu/en/geographical-distribution-2019-ncov-cases}{https://www.ecdc.europa.eu/en/geographical-distribution-2019-ncov-cases}. We remove, however, the data from the current day, since it can be updated by the ECDC. We assumed non-available data to be zero prior to the first case being recorded for each country. Whenever the daily recorded data was negative (reflecting previously confirmed cases being discarded), we replaced that information with a zero. This is the case for only nine out of 29,610 observations as of the 20th of May 2020.

According to the ECDC, the data on number of cases is based on reports from health authorities worldwide (up to 500 sources), which are screened by epidemiologists for validation prior to being recorded in the ECDC dataset.

\subsection*{Modelling framework}

We introduce a class of state-space hierarchical models for overdispersed count time series. Let $Y_{it}$ be the observed number of newly infected people at country $i$ and time $t$, with $i=1,\ldots,N$ and $t=1,\ldots,T$. We model $Y_{it}$ as a Negative binomial first-order Markov process, with
\begin{align*}
    Y_{it}|Y_{i,t-1} &\sim \mbox{NB}(\mu_{it},\psi)
\end{align*}
for $t=2,\ldots,T$. The parameterisation used here results in $\mbox{E}(Y_{it}|Y_{i,t-1})=\mu_{it}$ and $\mbox{Var}(Y_{it}|Y_{i,t-1})=\mu_{it}+\mu_{it}^2\psi$. The mean is modelled on the log scale as the sum of an autoregressive component ($\gamma_{it}$) and a component that accommodates outliers ($\Omega_{it}$), i.e.
\begin{align*}
    \log\mu_{it} &= \gamma_{it}+\Omega_{it}.
\end{align*}
To accommodate the temporal correlation in the series, the non-stationary autoregressive process $\left\lbrace \gamma_{it} \right\rbrace$ is set up as 
\begin{align}
    \gamma_{it} &= \phi_{it}\gamma_{it-1} + \eta_{it}, \mbox{ with } \eta_{it} \sim \mathrm{N}\left(0,\sigma^2_{\eta}\right),
    \label{eq:gamma_ar}
\end{align}
where $\eta_{it}$ is a Gaussian white noise process with mean $0$ and variance $\sigma_{\eta}^2$. Differently from standard AR(1)-type models, here $\phi_{it}$ is allowed to vary over time through an orthogonal polynomial regression linear predictor using the time as covariate, yielding
\begin{align}
    \phi_{it} &= \displaystyle \sum_{q=0}^{Q} (\beta_{q}+b_{iq})P_{q}(t), \mbox{ with } \bm{b}_{i} \sim \mathrm{N}_Q\left(\mathbf{0}, \Sigma_b\right)
    \label{eq:phi}
\end{align}
where $P_q(\cdot)$ is the function that produces the orthogonal polynomial of order $q$, with $P_0(x)=1$ for any real number $x$; $\beta_q$ are the regression coefficients and $\bm{b}_{i}$ is the vector of random effects, which are assumed to be normally distributed with mean vector $\mathbf{0}$ and variance-covariance matrix $\Sigma_b=\mathrm{diag}\left({\sigma^2_{b_0},\ldots,\sigma^2_{b_Q}}\right)$.

By assuming $\phi_{it}$ varying by country over time, we allow for a more flexible autocorrelation function. Iterating (\ref{eq:gamma_ar}) we obtain
\begin{align*}
    \displaystyle\gamma_{it} = \left(\prod_{k=2}^t\phi_{ik}\right)\gamma_{i1}+\sum_{j=2}^{t-1}\left[\left(\prod_{k=j+1}^t\phi_{ik}\right)\eta_{ij}\right]+\eta_{it}
\end{align*}
for $t = 3,\ldots,T$. Note that in the particular case where $\phi_{it}=\phi_{i}=\beta_{0}+b_{0i}$, then $\gamma_{it}= \phi_{i}^{t-1}\gamma_{i1}+\phi_{i}^{t-2}\eta_{i2}+\phi_{i}^{t-3}\eta_{i3}+ \ldots + \phi_{i}\eta_{it-1}+\eta_{it}$, which is equivalent to a country-specific AR(1) process. On the other hand, if $\phi_{it}=\phi_{i}=\beta_{0}$, then $\gamma_{it}= \phi^{t-1}\gamma_{i1}+\phi^{t-2}\eta_{i2}+\phi^{t-3}\eta_{i3}+ \ldots + \phi\eta_{it-1}+\eta_{it}$, which is equivalent to assuming the same autocorrelation parameter for all countries.

Finally, to accommodate extra-variability we introduce the observational-level random effect
\begin{align*}
    \Omega_{it} &= \lambda_{it}\omega_{it},
\end{align*}
where $\lambda_{it}\sim\mbox{Bernoulli}(\pi)$ and $\omega_{it}\sim\mbox{N}(0,\sigma^2_{\omega})$. When $\lambda_{it}=1$, then observation $y_{it}$ is considered to be an outlier, and the extra variability is modelled by $\sigma^2_{\omega}$. This can be seen as a mixture component that models the variance associated with outlying observations.

To forecast future observations $y_{i,t+1}^*$, we use the median of the posterior distribution of $Y_{i,t+1}|Y_{it}$. This is reasonable for short-term forecasting, since the error accumulates from one time step to the other. We produce forecasts for up to seven days ahead.

We fitted models considering different values for $Q$. Even though the results for $Q=3$ showed that all $\beta_q$ parameters were different from zero when looking at the 95\% credible intervals, we opted for $Q=2$ for the final model fit, since it improves convergence of the model, as well as avoids overfitting by assuming a large polynomial degree, while still providing the extra flexibility introduced by the autoregressive function~(\ref{eq:phi}). This can change, however, as more data becomes available for a large number of time steps.

\subsection*{Model validation}

We fitted the model without using the observations from the last seven days and obtained the forecasts $y_{it}^*$ for each country and day. We then compared the forecasts with the true observations $y_{it}$ for each day ahead, by looking initially at the Pearson correlation between them. We also computed the concordance correlation coefficient \cite{Lin1989, Oliveira2018a}, an agreement index that lies between $-1$ and $1$, given by
\begin{align*}
    \rho^{(CCC)}_t &= 1 -\frac{\mbox{E}\left[\left(Y_{t}^* - Y_{t}\right)^2\right]}{\sigma_{1}^{2}+\sigma_{2}^{2}+\left(\mu_{1}-\mu_{2}\right)^2}=\frac{2\sigma_{12}}{\sigma_{1}^{2}+\sigma_{2}^{2}+\left(\mu_{1}-\mu_{2}\right)^2}
\end{align*}
where $\mu_{1}=\mbox{E}\left(Y_{t}^*\right)$, $\mu_{2}=\mbox{E}\left(Y_{y}\right)$, $\sigma_{1}^{2}= \mbox{Var}\left(Y_{t}^*\right)$, $\sigma_{2}^{2}= \mbox{Var}\left(Y_{t}\right)$, and $\sigma_{12}=\mbox{Cov}\left(Y_{t}^*, Y_{2}\right)$. It can be shown that $\rho^{(CCC)}_t=\rho_t C_t$, where $\rho_t$ is the Pearson correlation coefficient (a measure of precision), and $C_t$ is the bias corrector factor (a measure of accuracy) at the $t-$th day ahead. $\rho_t$ measures how far each observation deviated from the best-fit line while $C_b \in (0,1]$ measures how far the best-fit line deviates from the identity line through the origin, defined as $C_{b}=2\left(v+v^{-1}+u^{2}\right)^{-1}$, where $v = \sigma^2_{1}/\sigma^2_{2}$ is a scale shift and $u = (\mu_{1} - \mu_{2}) / \sqrt{\sigma_1\sigma_2}$ is a location shift relative to the scale. When $C_b=1$ then there is no deviation from the identity line.

\subsection*{Model implementation}

The model is estimated using a Bayesian framework, and the prior distributions used are
\begin{align*}
    \boldsymbol\beta_{i} &\sim \mbox{N}_Q(\mathbf{0}, \mathbf{I}_Q\times1000) \\
    \sigma^{-2}_{b_q} &\sim \mbox{Gamma}(0.001, 0.001) \\
    \sigma^{-2}_{\eta} &\sim \mbox{Gamma}(0.001, 0.001) \\
    \sigma^{-2}_{\omega} &\sim \mbox{Gamma}(0.001, 0.001) \\
    \pi &\sim \mbox{Uniform}(0, 1)
\end{align*}

We used 3 MCMC chains, 2,000 adaptation iterations, 50,000 as burn-in, and 50,000 iterations per chain with a thinning of 25. We assessed convergence by looking at the trace plots, autocorrelation function plots and the Gelman-Rubin convergence diagnostic \cite{Gelman1992}.

All analyses were carried out using R software \cite{RcoreTeam2020} and JAGS \cite{Denwood2016}. Model fitting takes approximately seven hours to run in parallel computing, in a Dell Inspiron 17 7000 with 10th Generation Intel Core i7 processor, 1.80GHz$\times$8 processor speed, 16GB RAM plus 20GB of swap space, 64-bit integers, and the platform used is a Linux Mint 19.2 Cinnamon system version 5.2.2-050202-generic.

\subsection*{Clustering}

We used the last 60 values of the estimated autoregressive component to perform clustering so as to obtain sets of countries that presented a similar recent behaviour. First, we computed the dissimilarities between the estimated time series $\boldsymbol{\hat\gamma}_i$ between each pair of countries using the dynamic time warp (DTW) distance \cite{Muller2007,Montero2014}. Let $M$ be the set of all possible sequences of $m$ pairs preserving the order of observations in the form $r=((\hat\gamma_{i1}, \hat\gamma_{i^\prime 1}),\ldots,(\hat\gamma_{im}, \hat\gamma_{i^\prime m}))$. Dynamic time warping aims to minimise the distance between the coupled observations $(\hat\gamma_{it}, \hat\gamma_{i^\prime t})$. The DTW distance may be written as
\begin{align*}
d(\boldsymbol{\hat\gamma}_i,\boldsymbol{\hat\gamma}_{i^\prime})&=\min_{r\in M}\left(\sum_{t=1}^m|\hat\gamma_{it}-\hat\gamma_{i^\prime t}|\right).
\end{align*}
By using the DTW distance, we are able to recognise similar shapes in time series, even in the presence of shifting and/or scaling \cite{Montero2014}.

Then, we performed hierarchical clustering using the matrix of DTW distances using Ward's method, aimed at minimising the variability within clusters \cite{Murtagh2014}, and obtained ten clusters. Finally, we produced a dendrogram of the results of the hierarchical clustering analysis, with each cluster coloured differently so as to aid visualisation.

\subsection*{Code availability}

All code and datasets are made available as Supplementary Materials, and updated versions can also be downloaded from \href{https://github.com/Prof-ThiagoOliveira/covid_forecast}{https://github.com/Prof-ThiagoOliveira/covid\_forecast}.

We produced a Shiny app displaying the latest forecasts for the next seven days that is updated twice a week, and can be accessed at \href{https://prof-thiagooliveira.shinyapps.io/COVIDForecast}{https://prof-thiagooliveira.shinyapps.io/COVIDForecast}.

\bibliography{covid.bib}

\section*{Acknowledgements}

We would like to thank Prof. John Hinde for helpful comments when preparing this manuscript.

\section*{Author contributions statement}

T.P.O. and R.A.M. conceived and implemented the modelling framework, and wrote the manuscript. T.P.O. produced the Shiny visualisation app.

\section*{Additional information}

The authors declare no competing interests.

\section*{Supplementary Material}

\begin{figure}[ht!]
    \centering
    \includegraphics[width = \textwidth]{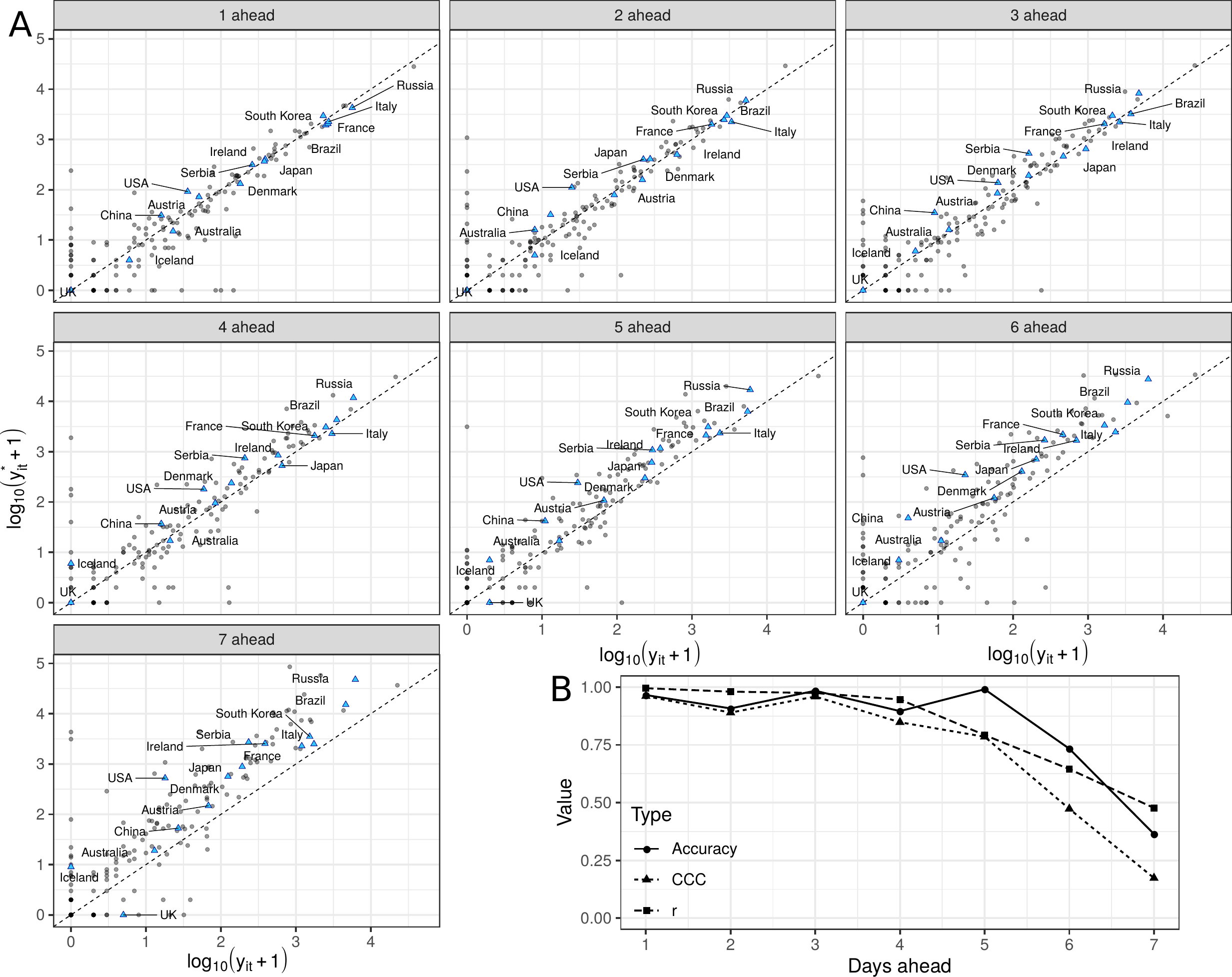}
    \caption{\textbf{A} Logarithm of the observed $y_{it}$ versus the forecasted daily number of cases $y^*_{it}$ for each country, for up to seven days ahead, where each day ahead constitutes one panel. The forecasts were obtained from the autoregressive state-space hierarchical negative binomial model, fitted using data up to 29-April-2020. The first day ahead corresponds to 30-April-2020, and the seventh to 06-May-2020. Each dot represents a country, and the sixteen countries shown in Figure~\ref{fig:gamma} are represented by blue triangles. We add 1 to the values before taking the logarithm. \textbf{B} Observed accuracy, concordance correlation coefficient (CCC) and Pearson correlation ($r$) between observed ($y_{it}$) and forecasted ($y^*_{it}$) values for each of the days ahead of 29-April-2020.}
    \label{fig:forecast1}
\end{figure}

\begin{figure}[ht!]
    \centering
    \includegraphics[width = \textwidth]{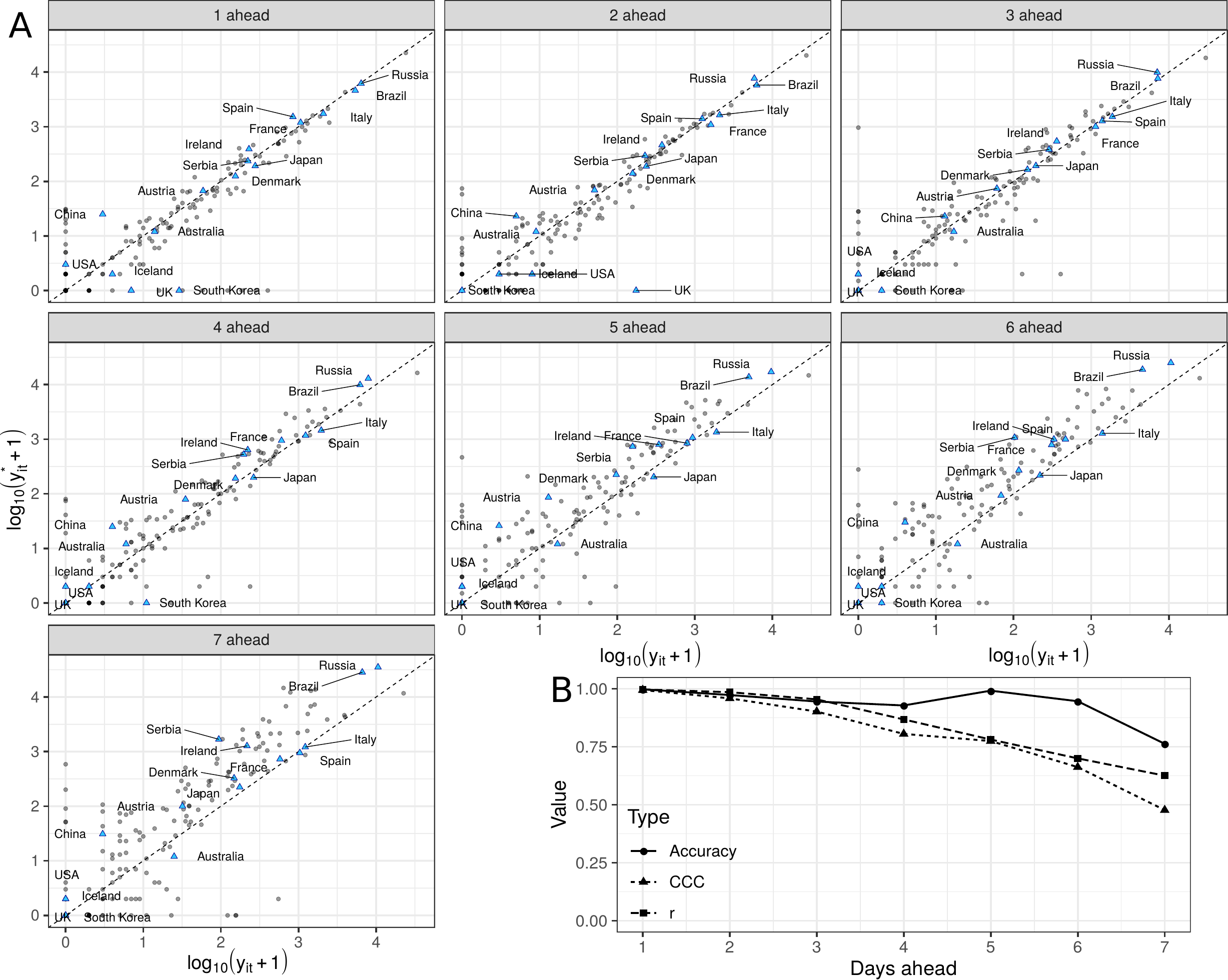}
    \caption{\textbf{A} Logarithm of the observed $y_{it}$ versus the forecasted daily number of cases $y^*_{it}$ for each country, for up to seven days ahead, where each day ahead constitutes one panel. The forecasts were obtained from the autoregressive state-space hierarchical negative binomial model, fitted using data up to 06-May-2020. The first day ahead corresponds to 07-May-2020, and the seventh to 13-May-2020. Each dot represents a country, and the sixteen countries shown in Figure~\ref{fig:gamma} are represented by blue triangles. We add 1 to the values before taking the logarithm. \textbf{B} Observed accuracy, concordance correlation coefficient (CCC) and Pearson correlation ($r$) between observed ($y_{it}$) and forecasted ($y^*_{it}$) values for each of the days ahead of 06-May-2020.}
    \label{fig:forecast2}
\end{figure}

\begin{figure}[ht!]
    \centering
    \includegraphics[width = \textwidth]{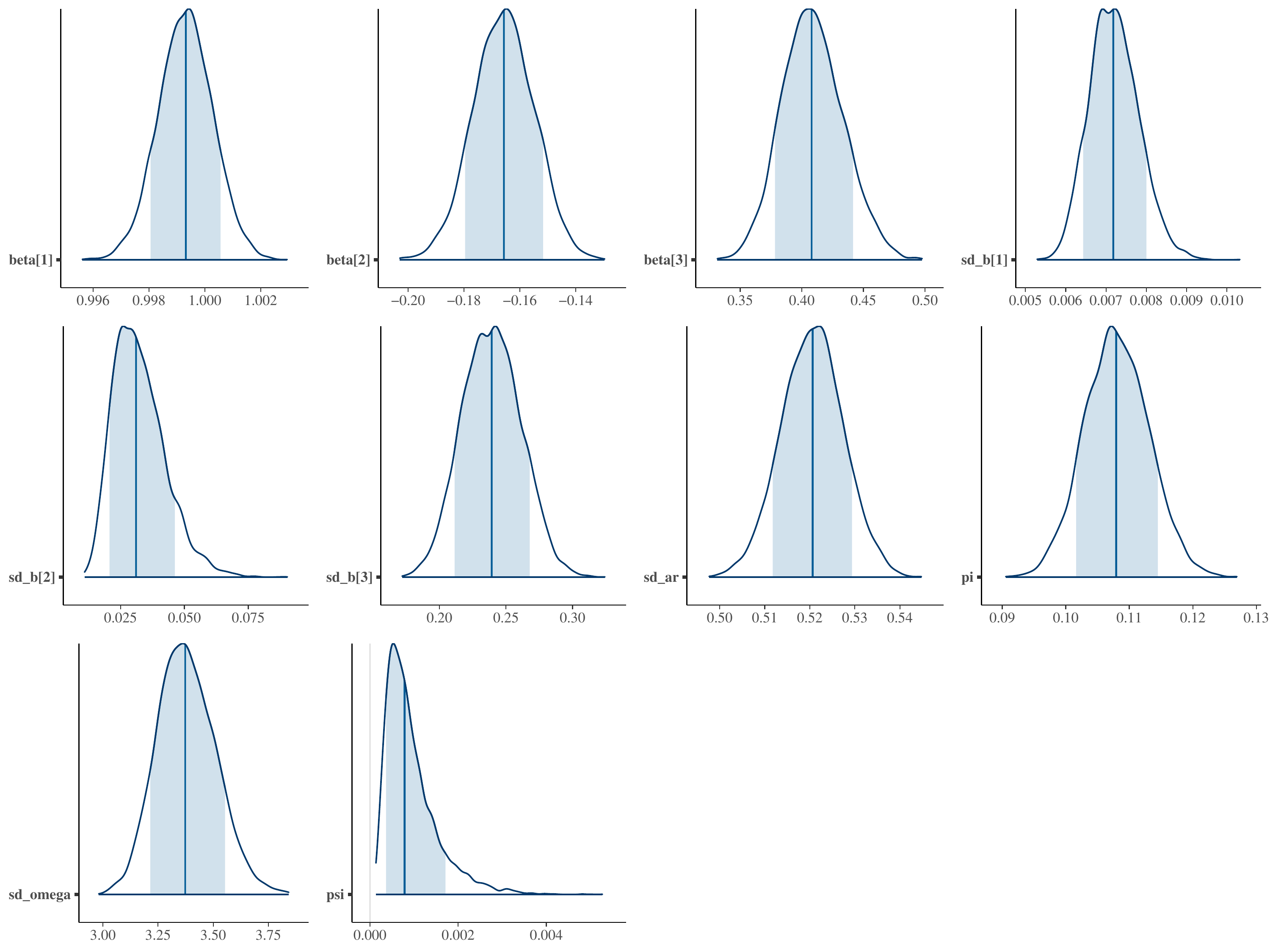}
    \caption{Density plots for the sampled posterior distributions for each model parameter.}
    \label{fig:mcmc}
\end{figure}

\begin{figure}[ht!]
    \centering
    \includegraphics[width = \textwidth]{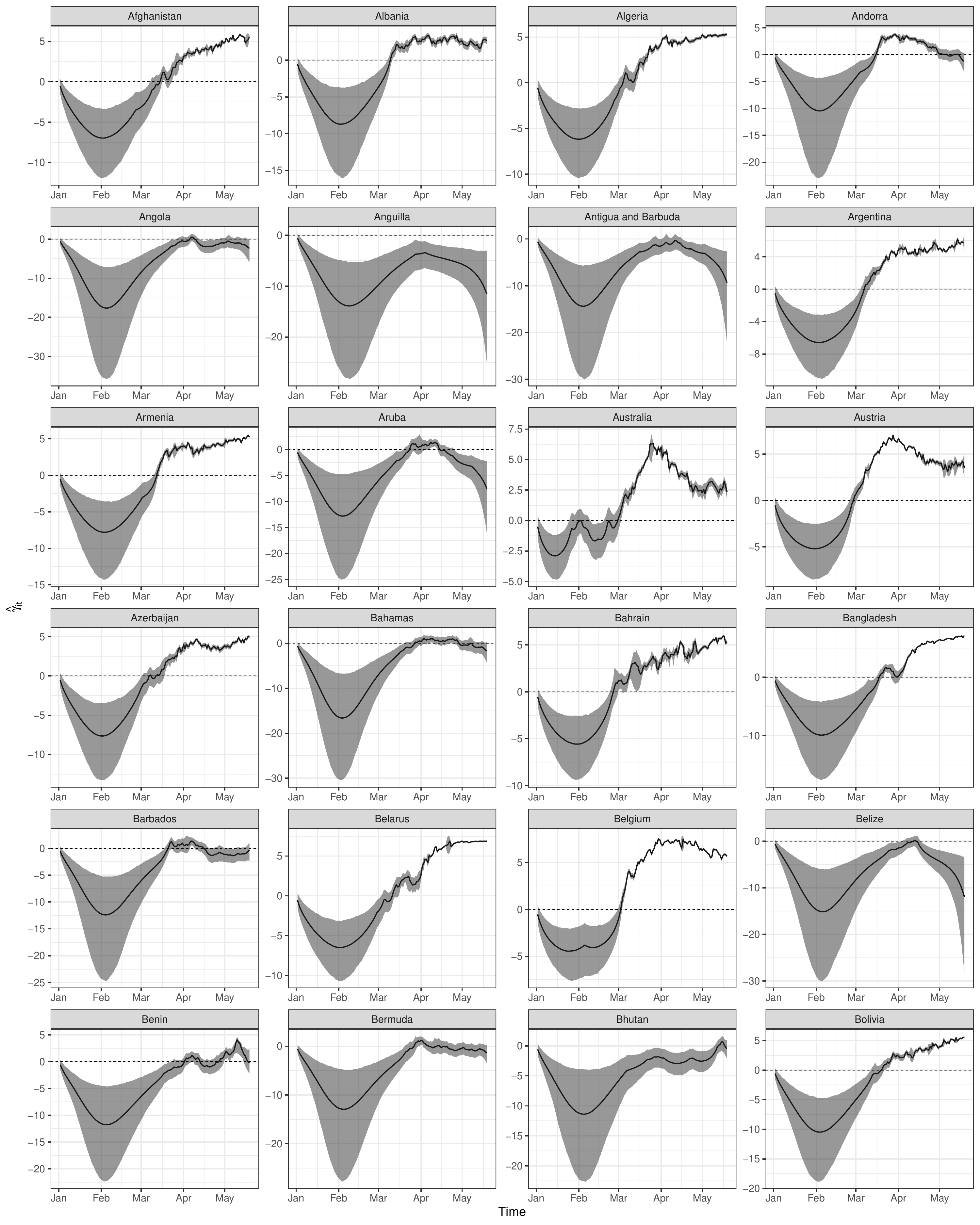}
    \caption{Posterior means of the autoregressive component $\gamma_{it}$ (solid lines) and associated 95\% credible intervals (shaded areas) for 24 countries from the pool of 210 countries and territories in the data, from 1-Jan-2020 until 20-May-2020.}
    \label{fig:gamma_suppl1}
\end{figure}

\begin{figure}[ht!]
    \centering
    \includegraphics[width = \textwidth]{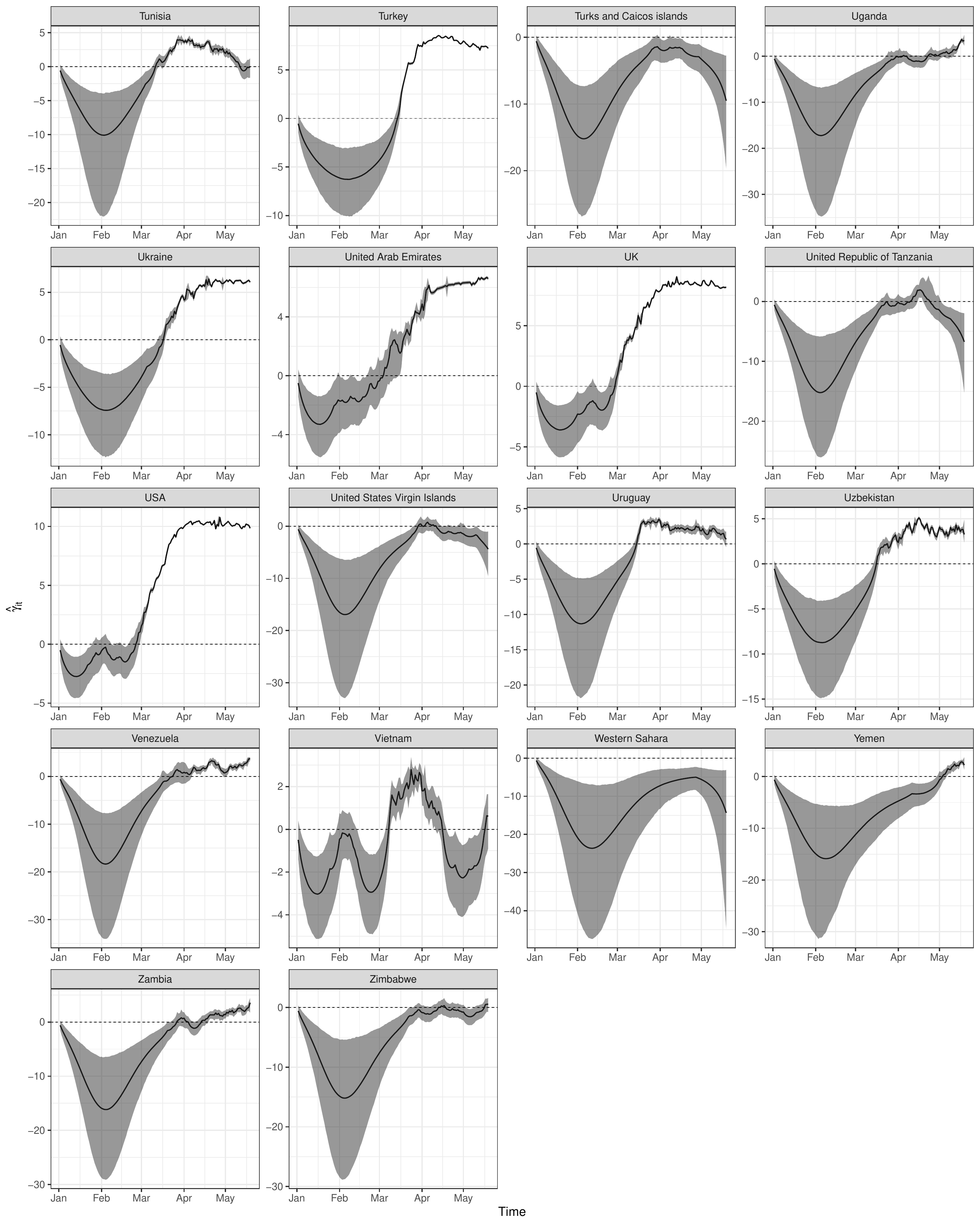}
    \caption{Posterior means of the autoregressive component $\gamma_{it}$ (solid lines) and associated 95\% credible intervals (shaded areas) for 18 countries from the pool of 210 countries and territories in the data, from 1-Jan-2020 until 20-May-2020.}
    \label{fig:gamma_suppl9}
\end{figure}

Table~\ref{tab:model} displays parameter estimates for the model fitted to the data up to 19-May-2020. Figure~\ref{fig:forecast1}--\ref{fig:forecast2} depicts the two additional forecast validation procedures we have carried out. Figure~\ref{fig:mcmc} displays density plots for the sampled posterior distributions for each model parameter. Figures~\ref{fig:gamma_suppl1}--\ref{fig:gamma_suppl9} display estimated autoregressive components $\hat\gamma_{it}$ for each country for the model fitted to the data up to 19-May-2020. 

\begin{table}[ht!]
    \centering
    \begin{tabular}{lrr} \hline
    Parameter & Estimate & $95\%$ CI [lower; upper] \\ \hline
    $\beta_{0}$  & $0.9993$ & $[0.9974, 1.0012]$ \\
    $\beta_{1}$  & $-0.1658$ & $[-0.1871, -0.1447]$ \\
    $\beta_{2}$  & $0.4090$ & $[0.3630, 0.4080]$\\
    $\sigma_{b_{0}}$  & $0.0072$ & $[0.0061, 0.0085]$ \\
    $\sigma_{b_{1}}$  & $0.0325$ & $[0.0168, 0.0565]$ \\
    $\sigma_{b_{2}}$  & $0.2392$ & $[0.1979, 0.2821]$ \\
    $\sigma_{\eta}$ & $0.5206$  & $[0.5071,0.5341]$ \\
    $\pi$ & $0.1080$ & $[0.0983,0.1180]$ \\
    $\sigma_{\omega}$ & $3.3797$ & $[3.1391, 3.6544]$ \\ 
    $\psi$ & $0.0009$ & $[0.0002,0.0025]$ \\ \hline
    \end{tabular}
    \caption{Parameter estimates and associated 95\% credible intervals (CI) for the fitted autoregressive hierarchical state-space negative binomial model. $\beta_0$, $\beta_1$ and $\beta_2$ are the fixed effects associated with the time orthogonal polynomials of order 0, 1 and 2, respectively; $\sigma_{b_0}$, $\sigma_{b_1}$ and $\sigma_{b_2}$ are the standard deviations of the random effects associated with the time orthogonal polynomials of order 0, 1 and 2, respectively; $\sigma_{\eta}$ is the standard deviation of the random autoregressive process; $\pi$ is the probability of an observation being an outlier and $\sigma_{\omega}$ is the standard deviation of the mixture component $\omega_{it}$; $\psi$ is the dispersion parameter of the negative binomial distribution.}
    \label{tab:model}
\end{table}

\begin{figure}[ht!]
    \centering
    \includegraphics[width = \textwidth]{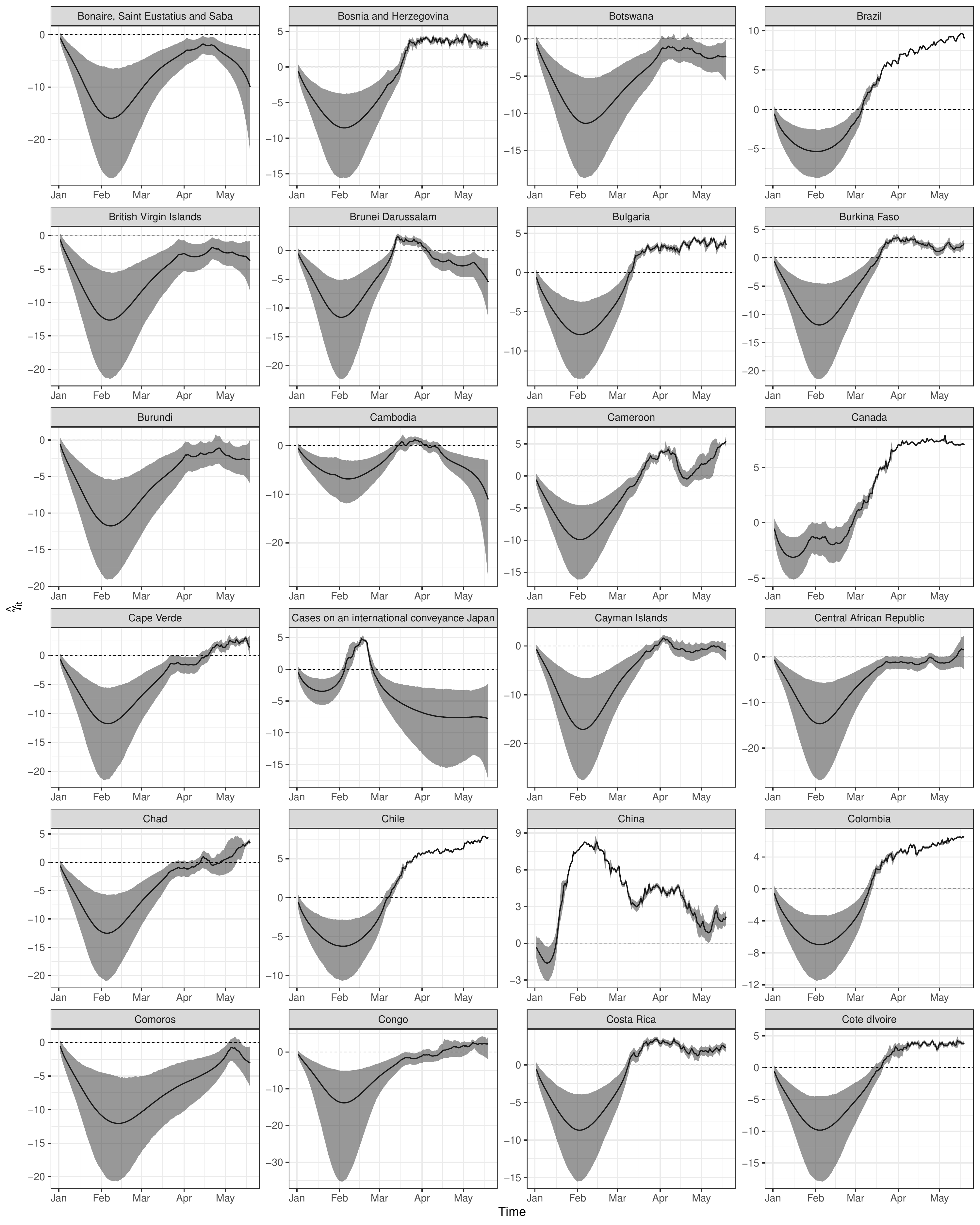}
    \caption{Posterior means of the autoregressive component $\gamma_{it}$ (solid lines) and associated 95\% credible intervals (shaded areas) for 24 countries from the pool of 210 countries and territories in the data, from 1-Jan-2020 until 20-May-2020.}
\end{figure}

\begin{figure}[ht!]
    \centering
    \includegraphics[width = \textwidth]{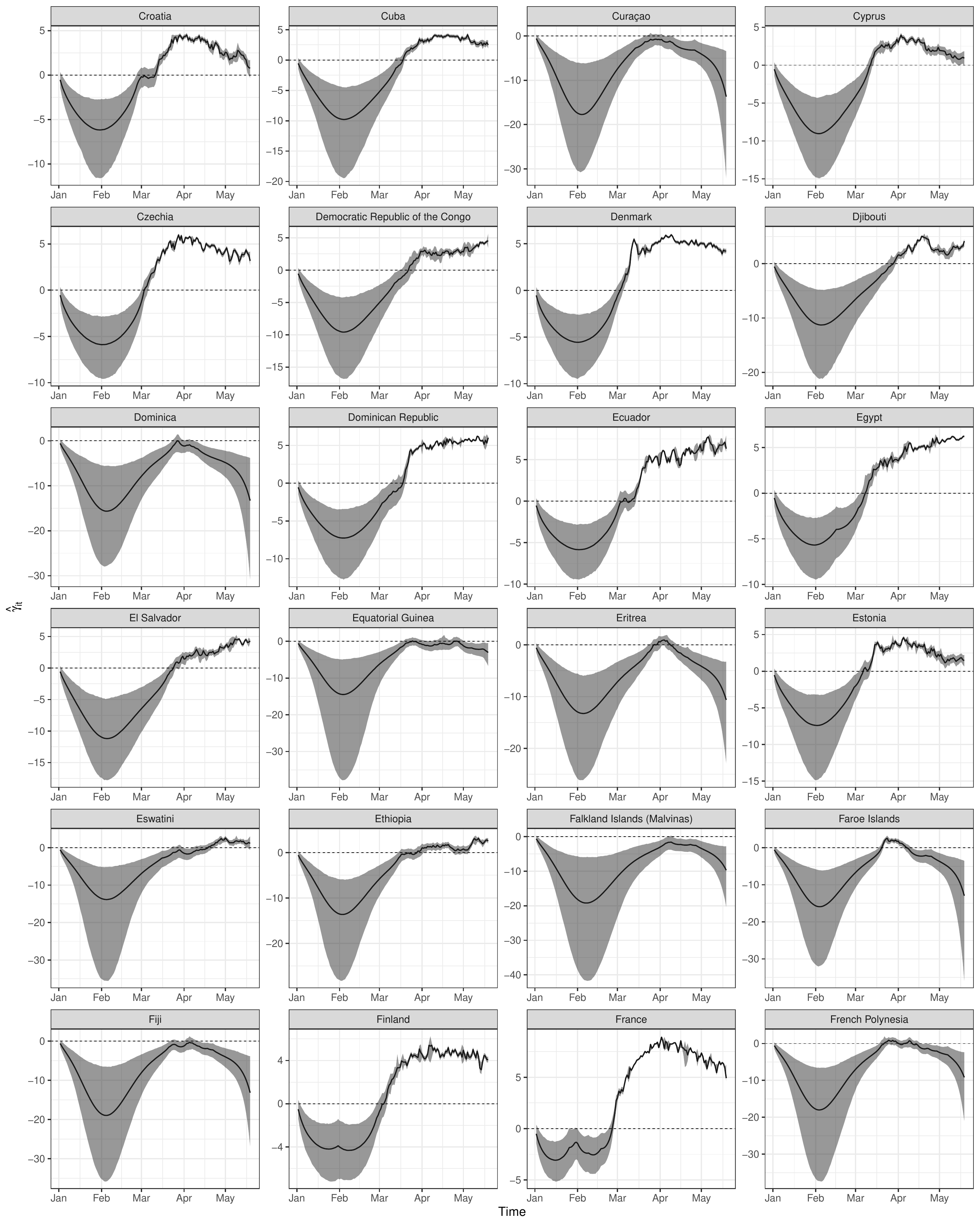}
    \caption{Posterior means of the autoregressive component $\gamma_{it}$ (solid lines) and associated 95\% credible intervals (shaded areas) for 24 countries from the pool of 210 countries and territories in the data, from 1-Jan-2020 until 20-May-2020.}
\end{figure}

\begin{figure}[ht!]
    \centering
    \includegraphics[width = \textwidth]{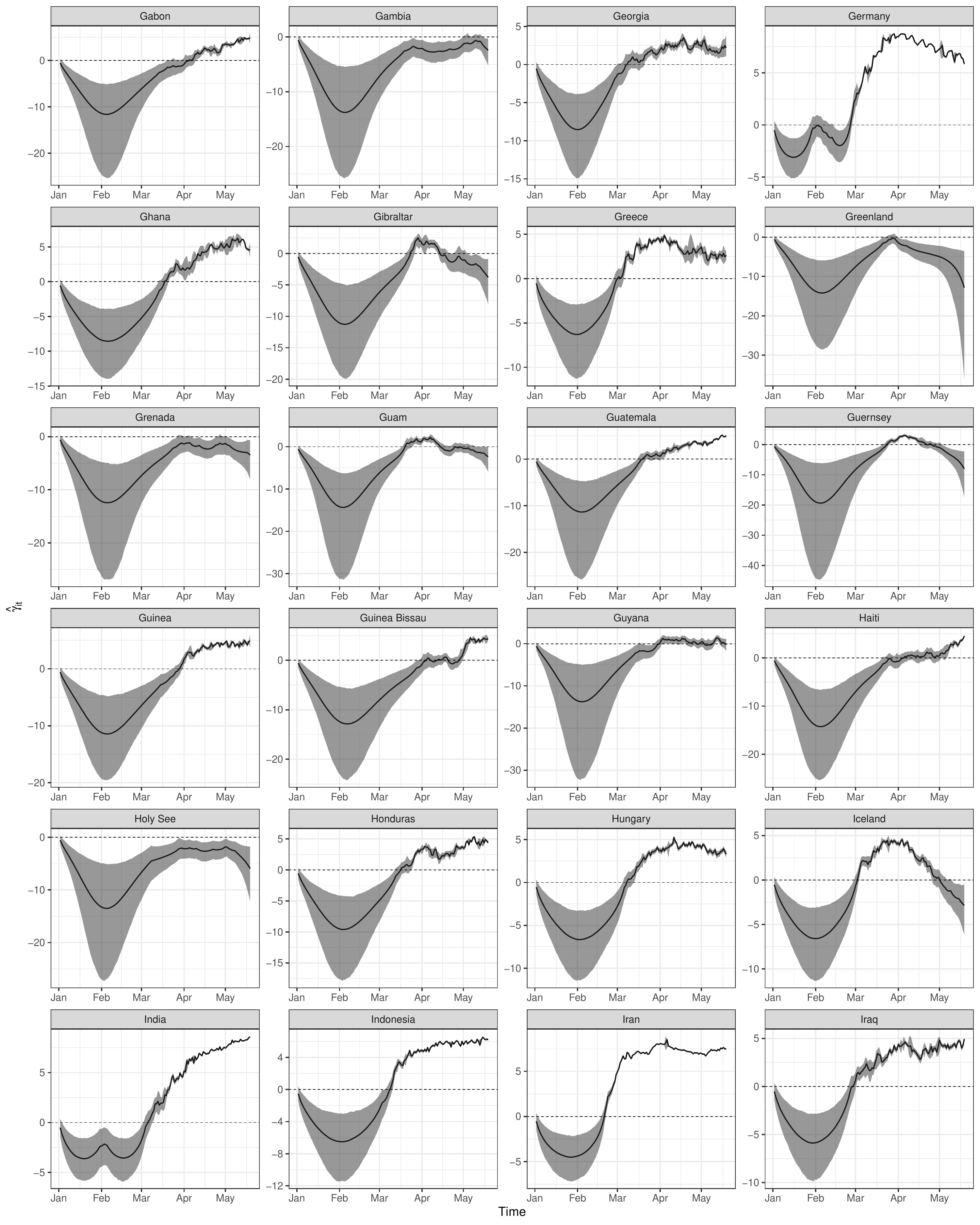}
    \caption{Posterior means of the autoregressive component $\gamma_{it}$ (solid lines) and associated 95\% credible intervals (shaded areas) for 24 countries from the pool of 210 countries and territories in the data, from 1-Jan-2020 until 20-May-2020.}
\end{figure}

\begin{figure}[ht!]
    \centering
    \includegraphics[width = \textwidth]{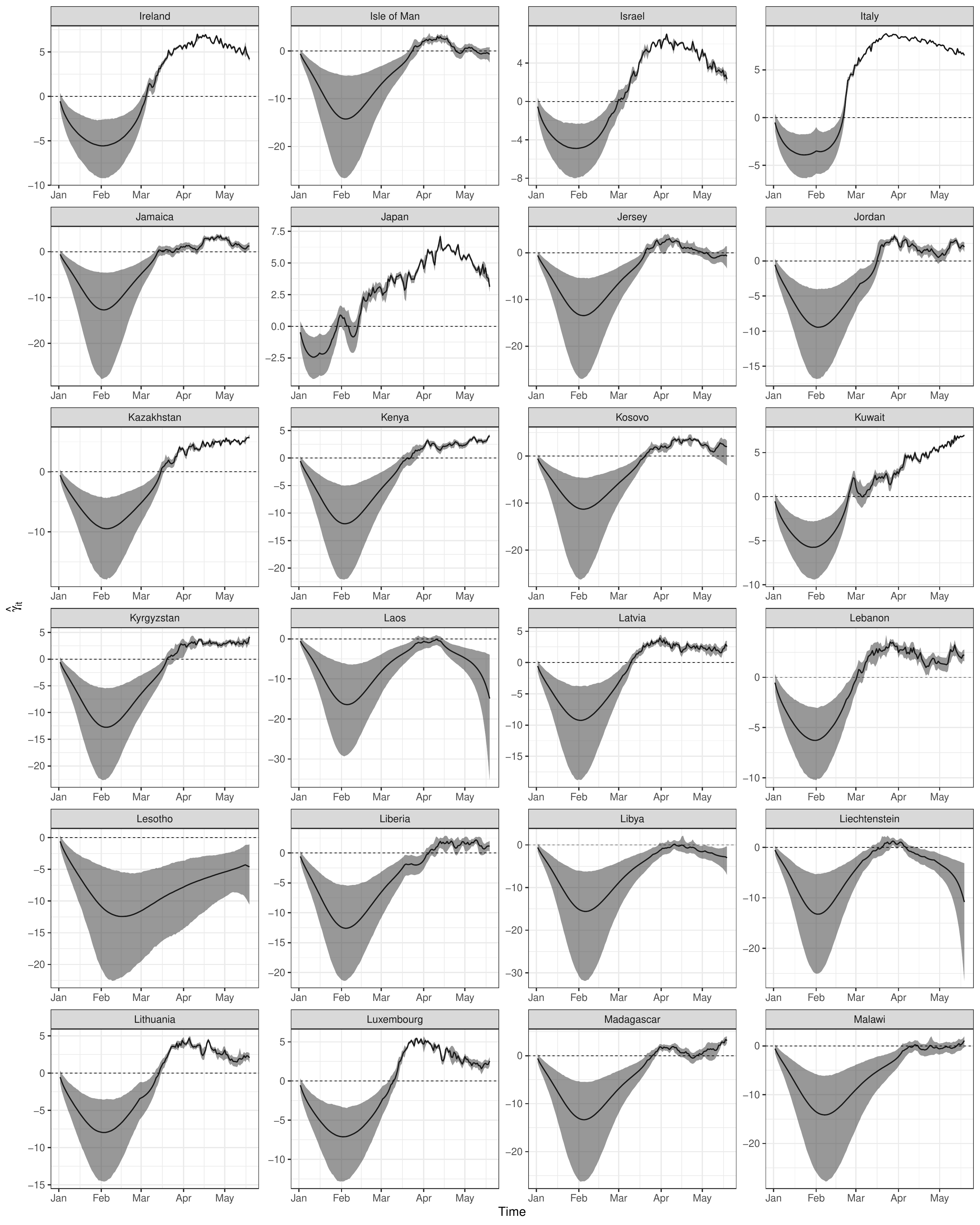}
    \caption{Posterior means of the autoregressive component $\gamma_{it}$ (solid lines) and associated 95\% credible intervals (shaded areas) for 24 countries from the pool of 210 countries and territories in the data, from 1-Jan-2020 until 20-May-2020.}
\end{figure}

\begin{figure}[ht!]
    \centering
    \includegraphics[width = \textwidth]{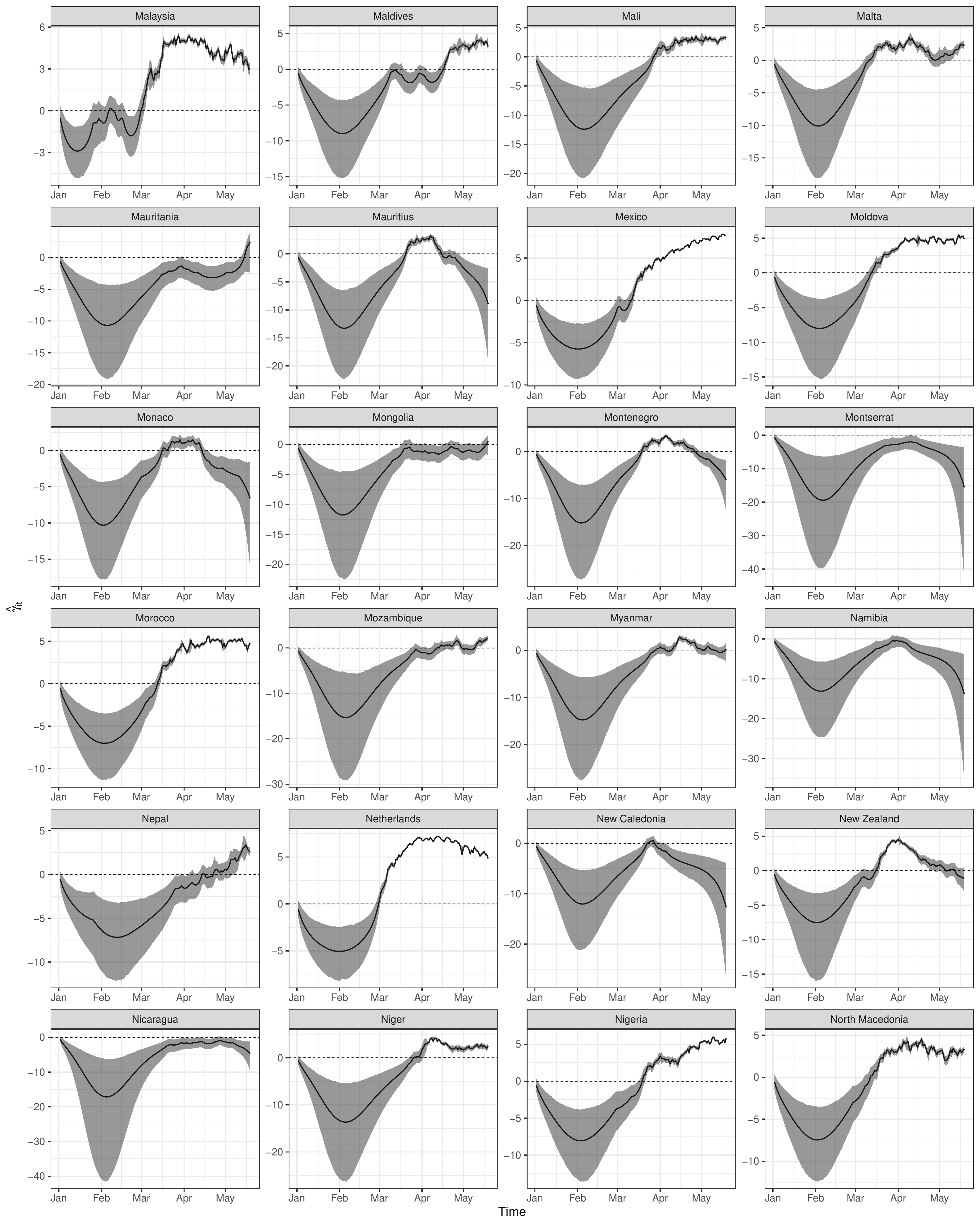}
    \caption{Posterior means of the autoregressive component $\gamma_{it}$ (solid lines) and associated 95\% credible intervals (shaded areas) for 24 countries from the pool of 210 countries and territories in the data, from 1-Jan-2020 until 20-May-2020.}
\end{figure}

\begin{figure}[ht!]
    \centering
    \includegraphics[width = \textwidth]{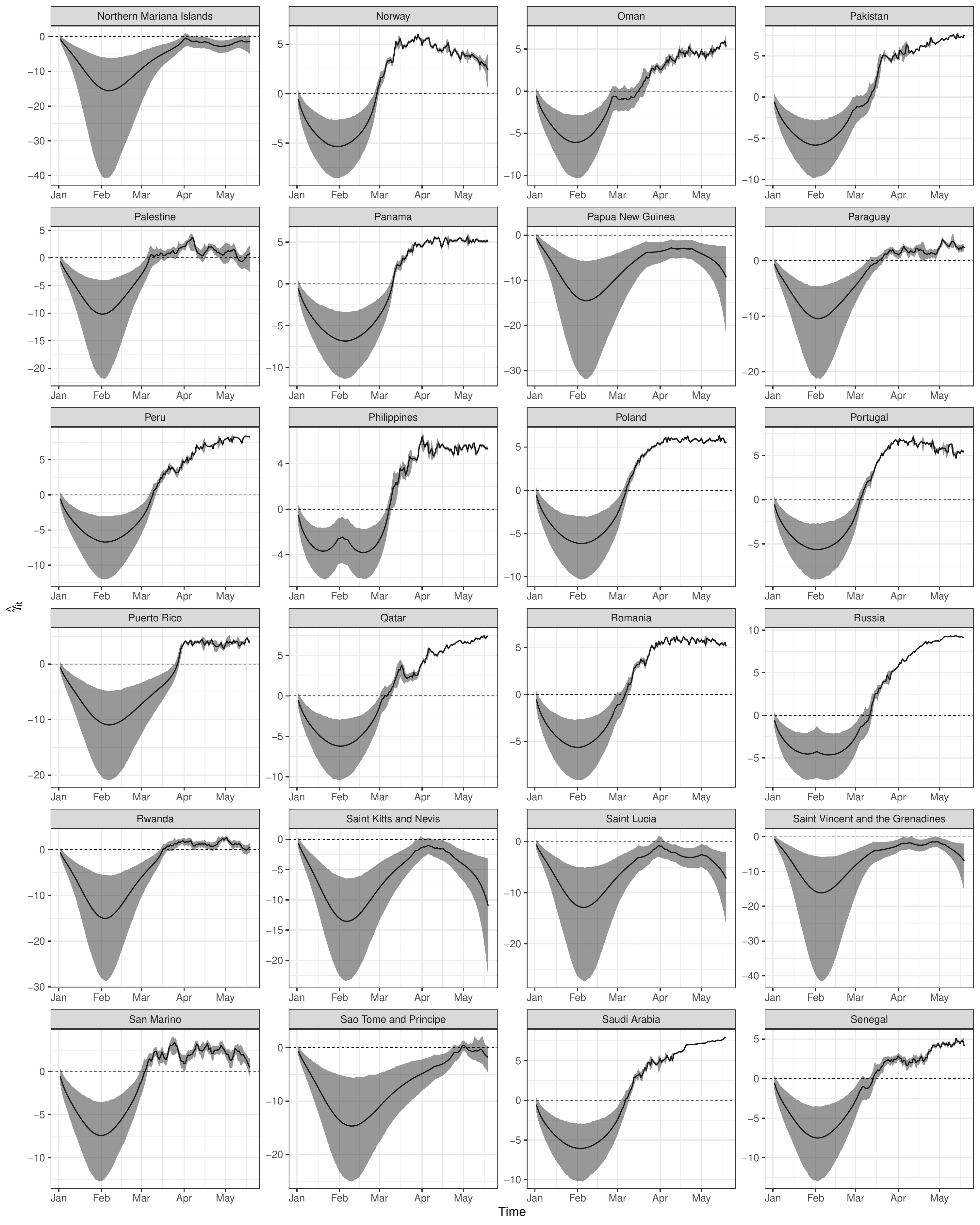}
    \caption{Posterior means of the autoregressive component $\gamma_{it}$ (solid lines) and associated 95\% credible intervals (shaded areas) for 24 countries from the pool of 210 countries and territories in the data, from 1-Jan-2020 until 20-May-2020.}
\end{figure}

\begin{figure}[ht!]
    \centering
    \includegraphics[width = \textwidth]{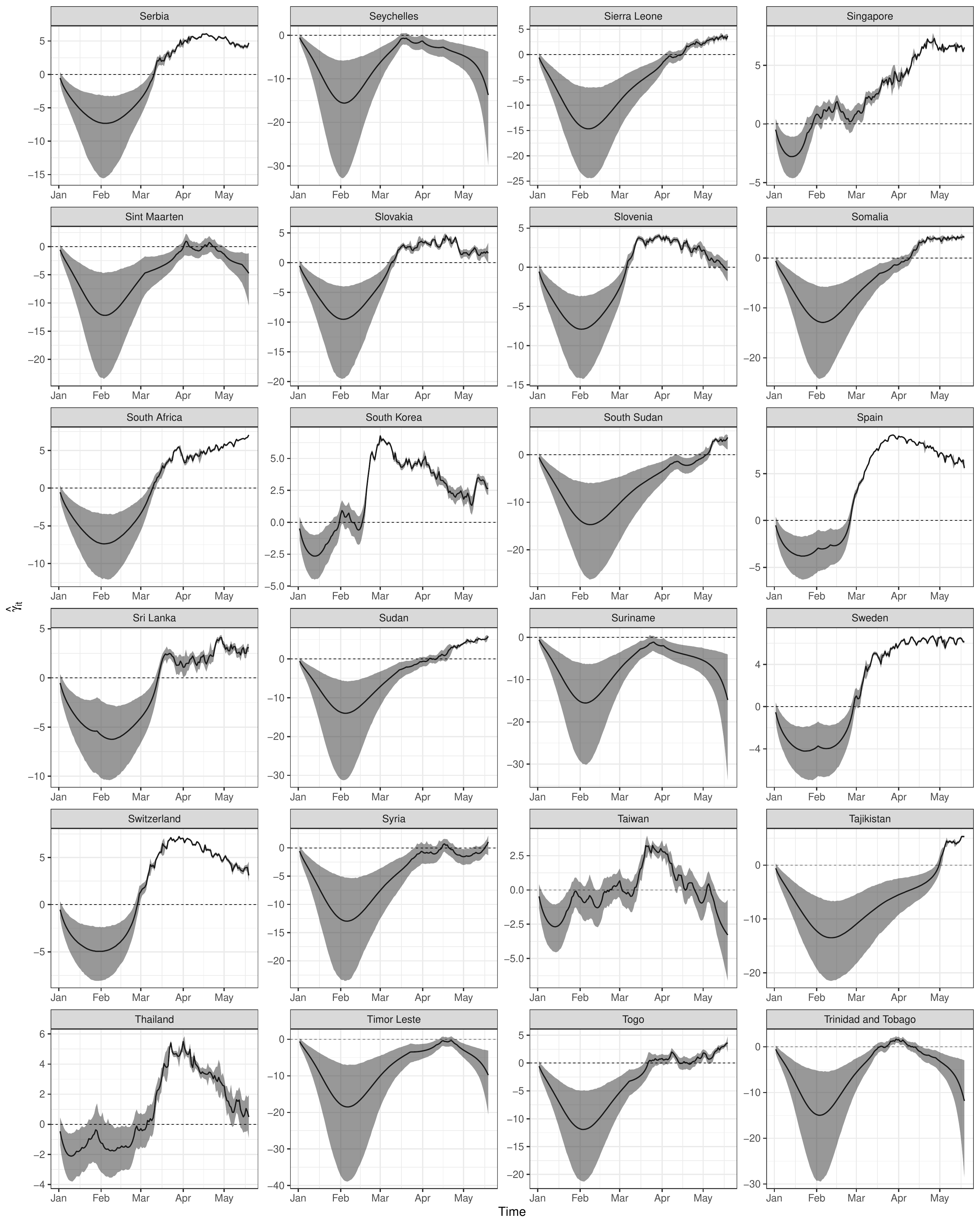}
    \caption{Posterior means of the autoregressive component $\gamma_{it}$ (solid lines) and associated 95\% credible intervals (shaded areas) for 24 countries from the pool of 210 countries and territories in the data, from 1-Jan-2020 until 20-May-2020.}
\end{figure}

\end{document}